\documentclass[twocolumn,tradiabstract]{aa} 

\def\bibfiles{biblio}
\def\aareferences{\bibliographystyle{aabib}
                 \bibliography{aajour,\bibfiles}}

\usepackage{natbib}
\usepackage{graphicx}
\usepackage{color}

\begin{document}

%
\title{High frequency waves in the corona due to null points}

\titlerunning{High frequency waves and null points}

\author{I.C. Santamaria\inst{1,2}, E. Khomenko\inst{1,3,4}, M. Collados\inst{1,3}, A. de Vicente\inst{1,3}}
\authorrunning{I.C. Santamaria et al.}

\institute{Instituto de Astrof\'{\i}sica de Canarias, 38205 La Laguna, Tenerife, Spain
\and Centre for mathematical Plasma Astrophysics, Department of Mathematics, KU Leuven, Celestijnenlaan 200B, B-3001 Leuven, Belgium
\and Departamento de Astrof\'{\i}sica, Universidad de La Laguna, 38205, La Laguna, Tenerife, Spain 
\and Main Astronomical Observatory, NAS, 03680, Kyiv, Ukraine}

\date{Received; Accepted }

\abstract {This work aims to understand the behavior of non-linear waves in the vicinity of a coronal null point. In previous works we have showed that high frequency waves are generated in such magnetic configuration. This paper studies those waves in detail in order to provide a plausible explanation of their generation. We demonstrate that slow magneto-acoustic shock waves generated in the chromosphere propagate through the null point and produce a train of secondary shocks that escape along the field lines. A particular combination of the shock wave speeds generates waves at a frequency of 80 mHz. We speculate that this frequency may be sensitive to the atmospheric parameters in the corona and therefore can be used to probe the structure of this solar layer.}


\maketitle

\section{Introduction}
Energy transport by MHD waves in the solar atmosphere is strongly influenced by the magnetic field topology. The highly dynamic solar magnetic fields lead to a wide variety of large and small scale magnetic structures. As a particular case, null points are prominent singularities located in the chromosphere and corona in which the magnetic field strength, and as a consequence the Alfv\'{e}n speed, drops to zero in extremely small spatial scales. Null points cannot be directly observed with current instrumentation for two main reasons. On the one hand, null points are very small scale features, requiring an extremely high spatial resolution for their detection. On the other hand, the measurement of coronal magnetic fields is still hard to achieve \citep{Lin+etal2004, VanDoorsselaere+etal2008c}. However, numerical simulations and extrapolations of phostospheric magnetic fields predict ubiquitous null points in the solar chromosphere and corona \citep{Galsgaard+Nordlund1997, Longcope2005}. 

Wave propagation in the presence of a null point changes drastically compared to its behavior in the absence of those discontinuities \citep{McLaughlin+etal2011, Santamaria+etal2015}. As the Alfv\'{e}n speed approaches to a zero value, pure Alfv\'{e}n waves cannot cross null points and are guided outwards along field lines. Additionally, fast magnetic-like waves are refracted around null points due to the strong gradient in the Alfv\'{e}n speed \citep{Nakariakov+Roberts1995, McLaughlin+Hood2004, Santamaria+etal2015}. Therefore, the only waves that can physically cross the null points are the acoustic-like slow magneto-acoustic waves. \citet{Santamaria+etal2015} modeled numerically the wave behavior around a coronal null point in the linear regime in which non-linearities do not develop. In their work these authors show how slow waves are trapped and resent outwards again in all directions by the influence of the magnetic topology of the null point. This phenomenon gives rise to a superposition of waves, with an amplitude decrease in certain locations.
In their most recent study, Santamaria et al. (2016) show how non-linearities change considerably the wave propagation around the null point and find that, due to the non-linear effects, high frequency waves of 80 mHz are generated. 

Recent works demonstrate that, although null points are extremely small scale magnetic field singularities, their implication in the solar dynamism is of great importance. They may drive highly energetic phenomena and contribute to chromospheric and coronal heating. In their study, \citet{McLaughlin+Hood2004} show that there is an accumulation of currents at null points, resulting in an increasing importance of dissipative processes that can convert the waves energy into heat. Additionally, null points may lead to reconnection events which also dissipate energy into heat \citep{McLaughlin+Hood2004b, McLaughlin+etal2009}. Furthermore, the magnetic reconnection events can drive strong and energetic phenomena in the solar corona, such as, solar flares \citep{Shibata+Magara2011}.

The present work, which is a continuation of \citet{Santamaria+etal2016a}, aims to give an explanation to the appearance of high frequency waves around the null point. We analyze the interaction between slow magneto-acoustic shock waves and a null point located in the corona. Such interaction generates "secondary" shock waves, leading to a jet-like phenomenon. This jet is strongly ejected outwards the null point along the field lines and is the responsible for the high frequencies around it.

\begin{figure}
\label{fig:mag_fields}
\includegraphics[width=9cm]{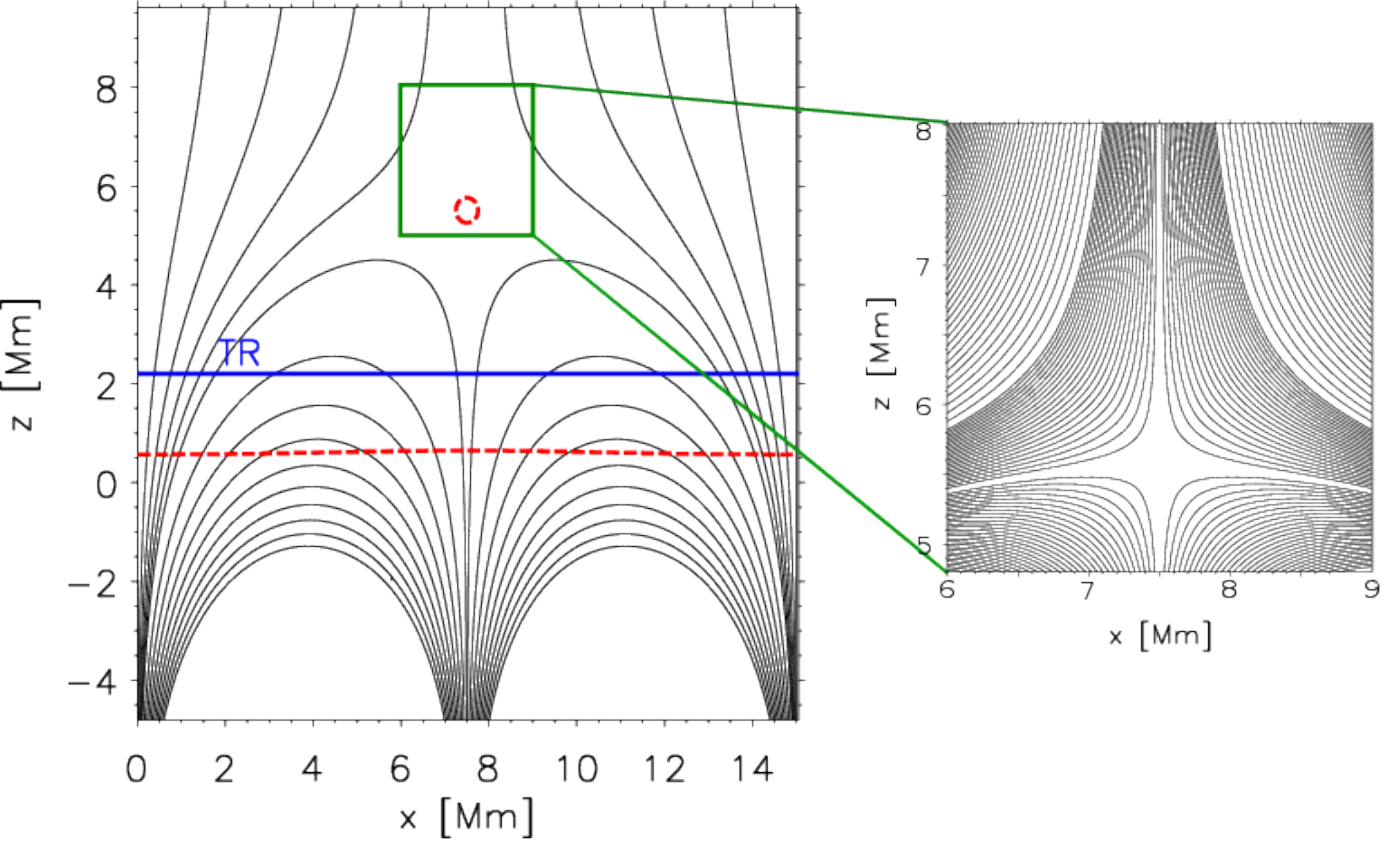}
\caption{Magnetic field lines in the entire simulation domain ($z=[-5,10] Mm$) (left) and in the enhanced region around the null point (right). The red dashed curves in the left-hand panel show the equipartition layers and the horizontal blue line marks the location of the transition region. }
\end{figure}
\section{Description of the simulations}
\label{sec:num_method}
The ideal two-dimensional MHD equations are numerically solved by means of the {\sc mancha} code \citep[see][for details]{Khomenko+Collados2006, Felipe+etal2010a}. The hydrostatic background model is plane parallel and vertically stratified from -5 Mm to 10 Mm, with the zero-height corresponding to that of the quiet Sun's photosphere in the VALC model \citep[for details see][]{Santamaria+etal2015, Santamaria+etal2016a}. The grid is homogeneous with a spatial resolution of $\Delta x = 50$ km and $\Delta z=10$ km in the horizontal and vertical directions, respectively. The magnetic field structure is potential and is composed by two vertical flux tubes of the same polarity connected by an arcade-shaped magnetic field. This magnetic field topology leads to a null point in the corona (see the field lines in the left-hand panel of Fig. \ref{fig:mag_fields}). The photospheric and coronal field strengths are 100 G and 10 G, respectively. After subtracting the equilibrium atmosphere, the MHD equations are solved for the perturbed variables.

The horizontal boundary conditions are chosen to be periodic and two Perfectly Matched Layers \citep[PML,][]{Berenger1996} have been implemented at the bottom and top boundaries. The PMLs are quasi-open boundary layers that absorb waves in order to avoid non-desired reflections. The energy injected in the simulation box can leave it through the open boundaries and therefore the total energy of the box is not expected to be conserved.

Waves are generated by perturbing the atmosphere with an instantaneous pressure pulse at the central foot point of the arcade \citep{Santamaria+etal2016a}. During its evolution, the pulse expands until it reaches the equipartition layer, located in the upper photosphere, where waves are partially converted and transmitted. Some of the waves continue propagating upwards to the transition region. At this location, most of the magnetic waves are refracted downwards due to the gradient in the Alfv\'{e}n speed and acoustic waves are partially reflected and partially transmitted into the corona. The slow waves propagating inside the vertical flux tubes reach higher coronal layers with almost no obstacles after crossing the transition region. On the contrary, the wave fronts propagating close to the null point on their way to the upper corona interact with the null point. Such interaction results in a high frequency jet-like phenomenon propagating outwards the null point.

\section{Wave behavior}
\label{sec:waves_beh}
\citet{Santamaria+etal2015} have shown that, in the linear regime, the null point traps waves and re-sends them again outwards in all directions. In the non-linear regime, the phenomenon that takes place in the vicinity of the null point is more intricate. The most remarkable difference between the linear and non-linear regimes at the vicinity of the null point is the generation of a jet-like phenomenon that is strongly driven outwards the null point. In order to perform an in-depth analysis of this phenomenon, we focus on the region close to the null point where the behavior of waves and magnetic field lines can be studied in detail (see the right panel of Figure \ref{fig:mag_fields}).

\begin{figure*}
\centering
\includegraphics[width=12cm]{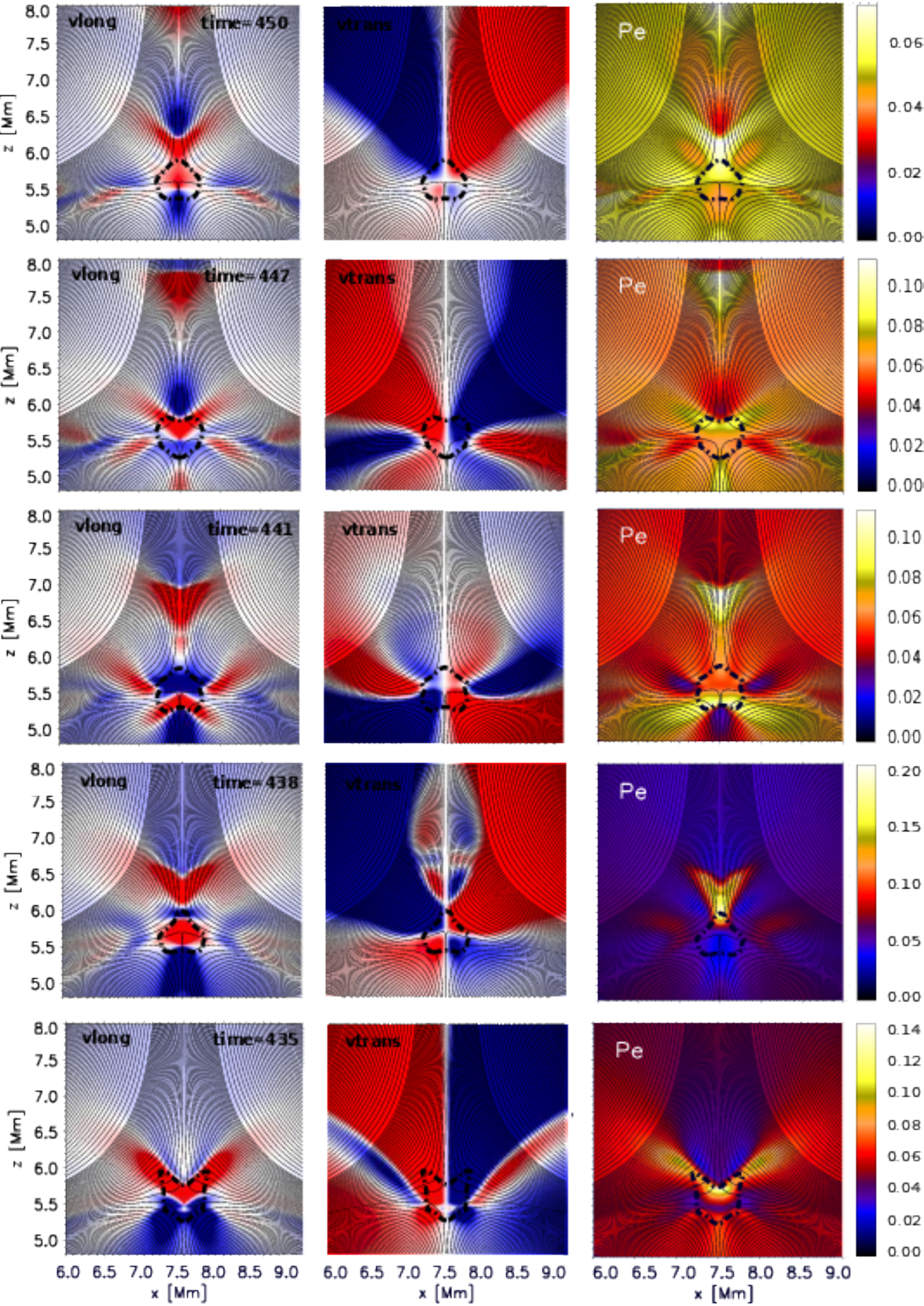}
\caption{Longitudinal (left) and transverse (middle) velocities multiplied by $\sqrt{\rho_{0}c_{s0}}$ and $\sqrt{\rho_{0}v_{A0}}$, respectively, in a region around the null point (a part of the simulation domain). The dashed closed curve is the contour where $\beta=1$ and solid lines are magnetic field lines. Right: fluctuations of the total gas pressure, i.e. the sum of the equilibrium and perturbed gas pressure, in the same region around the null point. Units are [dyn/cm$^{2}$]. Time increases from bottom to top. For a better visualization a movie of the propagation is available.}
\label{fig:vlvt_nullpoint}
\end{figure*}

Figure \ref{fig:vlvt_nullpoint} shows the evolution of acoustic and magnetic energy flux proxies, as well as the total gas pressure fluctuations.\footnote{Note that {\sc mancha} code solves the equations for perturbations. By total gas pressure, we mean the sum of the equilibrium and perturbed gas pressures.}. The figure covers the 15 seconds of the simulation in which the waves are most prominent and whose duration is adequate in comparison with the high-frequency waves period. 
Focusing on the left hand panel of the figure (longitudinal acoustic energy flux proxy), it can be seen how small-scale wave fronts (in red) propagate outwards from the null point. During the time interval from the first snapshot (bottom panels) to the last one (top panels), 15 seconds later, these wave fronts experience a displacement of about 2.5 Mm. Thus the wave front propagates with a velocity of about 170 km/s, close to the sound speed around the null point. 15 seconds after the appearance of the first front, a second front is being generated (at $t=450$ seconds).
Such features could, in principle, be produced by magnetic reconnection events and the corresponding energy release. Alternatively, at the null point pressure pulses may be generated given that the plasma $\beta$ around the null point is larger than one. The gas pressure dominates at the null point and around it, and hence, compressible waves can be produced, propagating outwards from it as the result of a piston effect.

One of the main conditions to be satisfied for reconnection to occur is that the magnetic Reynold's number has to be much lower than one ($R_{m}<<1$) at the reconnection site, making the Ohmic diffusion term large compared to the convective term in the induction equation. Our simulations provide us all the necessary information to calculate the magnetic Reynold's Number, $R_{m}$, from the parameters of the model in order to verify this condition. 
\begin{equation} \label{eq:Rm}
R_{m}=\frac{\vert \nabla \times (\textbf{v} \times \textbf{B}) \vert}{\vert \nabla \times (\eta \textbf{J})\vert} 
\end{equation}

In the {\sc mancha} code, used to solve the MHD equations, the Ohmic diffusion term in the induction equation is replaced by its artificial analog, and is given in terms of numerical hyperdiffusion \citep[as introduced by][]{Nordlund+Stein1990}. Simulations can resolve variations of the physical parameters down to scales imposed by the numerical resolution of the simulation grid (10 km and 50 km in the vertical and horizontal directions, respectively). The physical Ohmic diffusion term is too small at such spatial scales to produce reconnection on the observed short time scales. The substitution of the Ohmic diffusion by a numerical analog term allows to pick the details of the reconnection process, but not its physical origin, similar to the other numerical codes \citep{Vogler+etal2005, Gudiksen+etal2011}. Therefore, we performed the calculation of $R_m$, as given by Eq. \ref{eq:Rm}, using the diffusivity, $\eta$, used in the code. The result is that during the time interval in which the jet-like phenomenon is developed, $R_{m}>>$1. This means that there is no diffusion region in the immediate surroundings of the null point, and therefore, magnetic reconnection cannot take place.
\begin{figure}
\centering
\includegraphics[width=7cm]{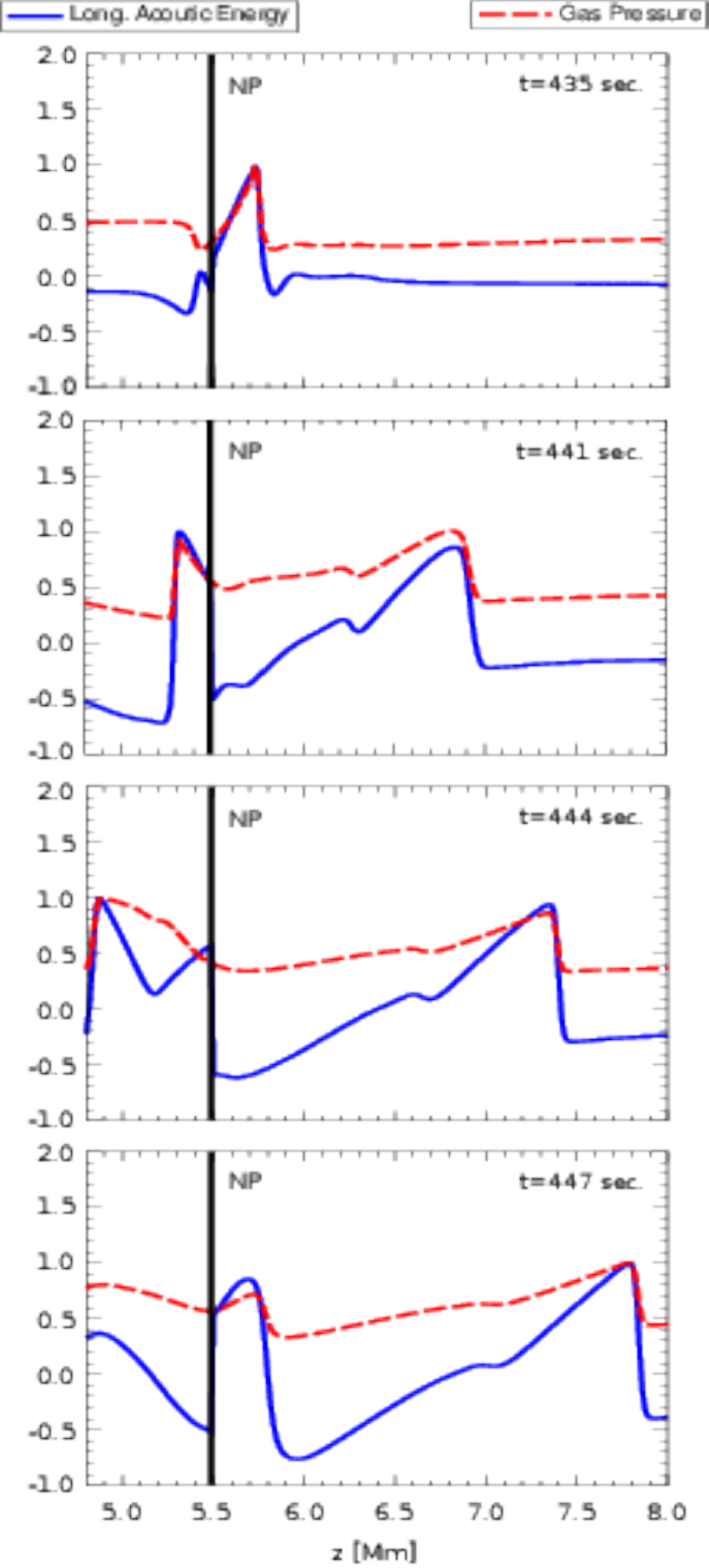}
\caption{Vertical cuts at the location of the null point showing the time evolution (from top to bottom) of the acoustic energy flux proxy (blue) and total gas pressure flutuations (red dashed line). Both variables have been normalized to their maximum values to better compare them. The vertical black line marks the location of the null point.} \label{fig:plot_pres_fl}
\end{figure}

Consider now the second alternative, i.e. the total gas pressure fluctuations around the null point. Figure \ref{fig:vlvt_nullpoint} shows the longitudinal acoustic and transverse magnetic energy proxies (left hand and middle panels) and the total gas pressure fluctuations (right hand panel) in the small region around the null point shown in Figure \ref{fig:mag_fields}. The acoustic energy proxy shows strongly driven waves coming out from the null point to the upper corona (left panels), while the magnetic energy flux proxy (middle panels) shows a very different evolution. The magnetic wave fronts are refracted around the null point due to the strong gradients in the Alfv\'{e}n speed. This behavior makes evident the acoustic nature of the jet-like feature being studied. Simultaneously, the evolution of the gas pressure (right panels) also manifests a jet-like feature, showing that both acoustic energy and gas pressure fluctuations have very similar patterns (see the provided movie attached to Fig. \ref{fig:vlvt_nullpoint}). For a clear visualization of those similarities in the patterns, Figure \ref{fig:plot_pres_fl} shows vertical cuts at the location of the null point for both the gas pressure and acoustic energy flux proxy variations. It can be clearly noticed that for all time steps the shape and position of the perturbations are similar, and hence, we conclude that the whole phenomenon is pressure driven.

In order to understand the origin of the gas pressure enhancement that drives the acoustic waves outwards from the null point, we have first checked if there are any slow magneto-acoustic shock waves propagating in the vicinity of the null point during the time interval. 
\begin{figure}
\centering
\includegraphics[width=6cm]{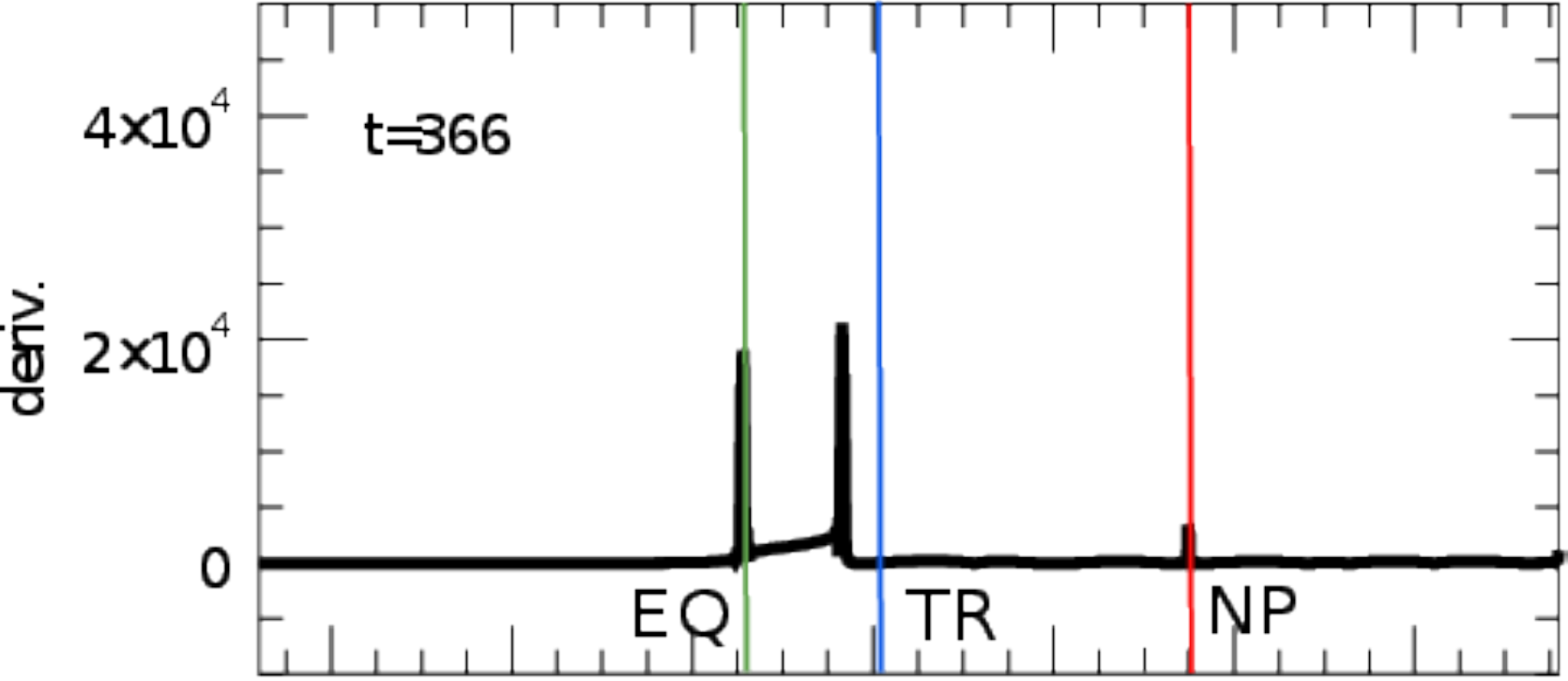}
\\
\includegraphics[width=6cm]{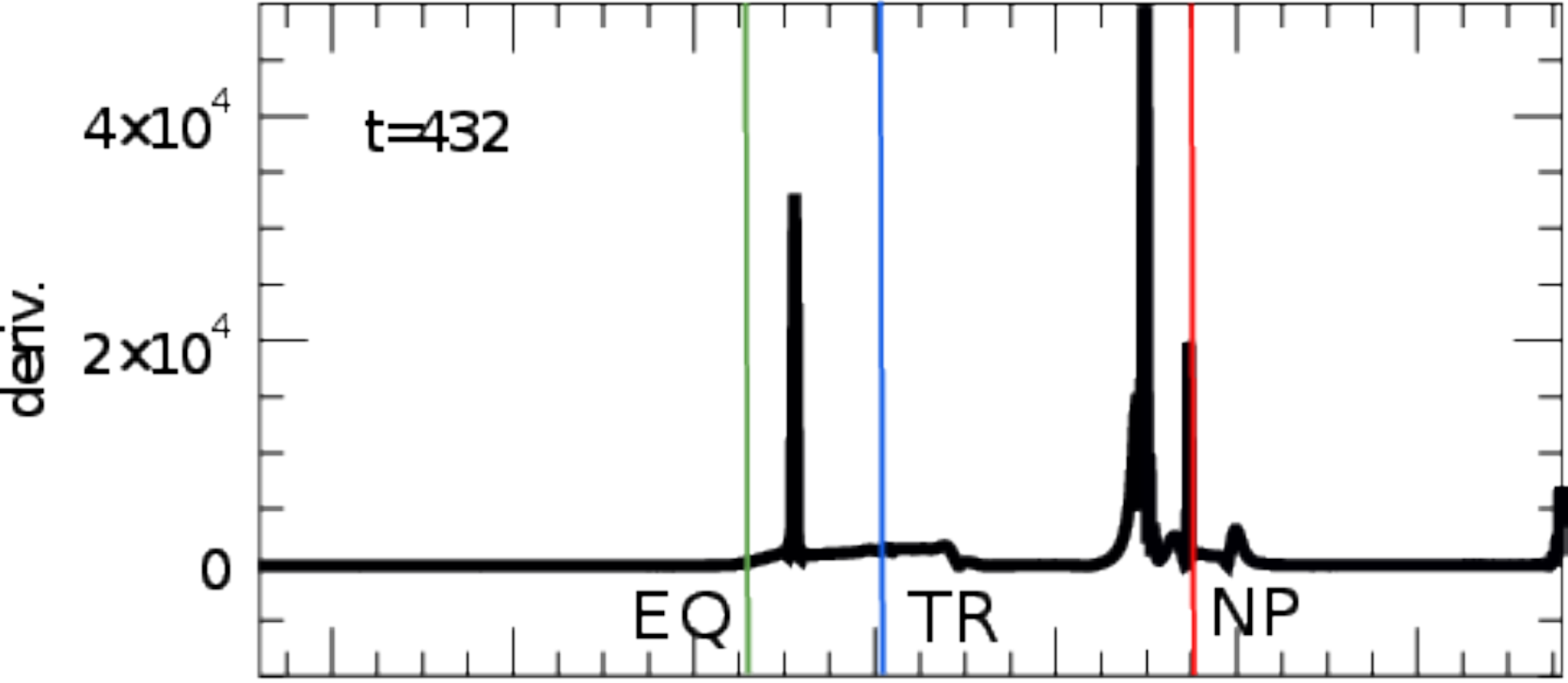}
\\
\includegraphics[width=6cm]{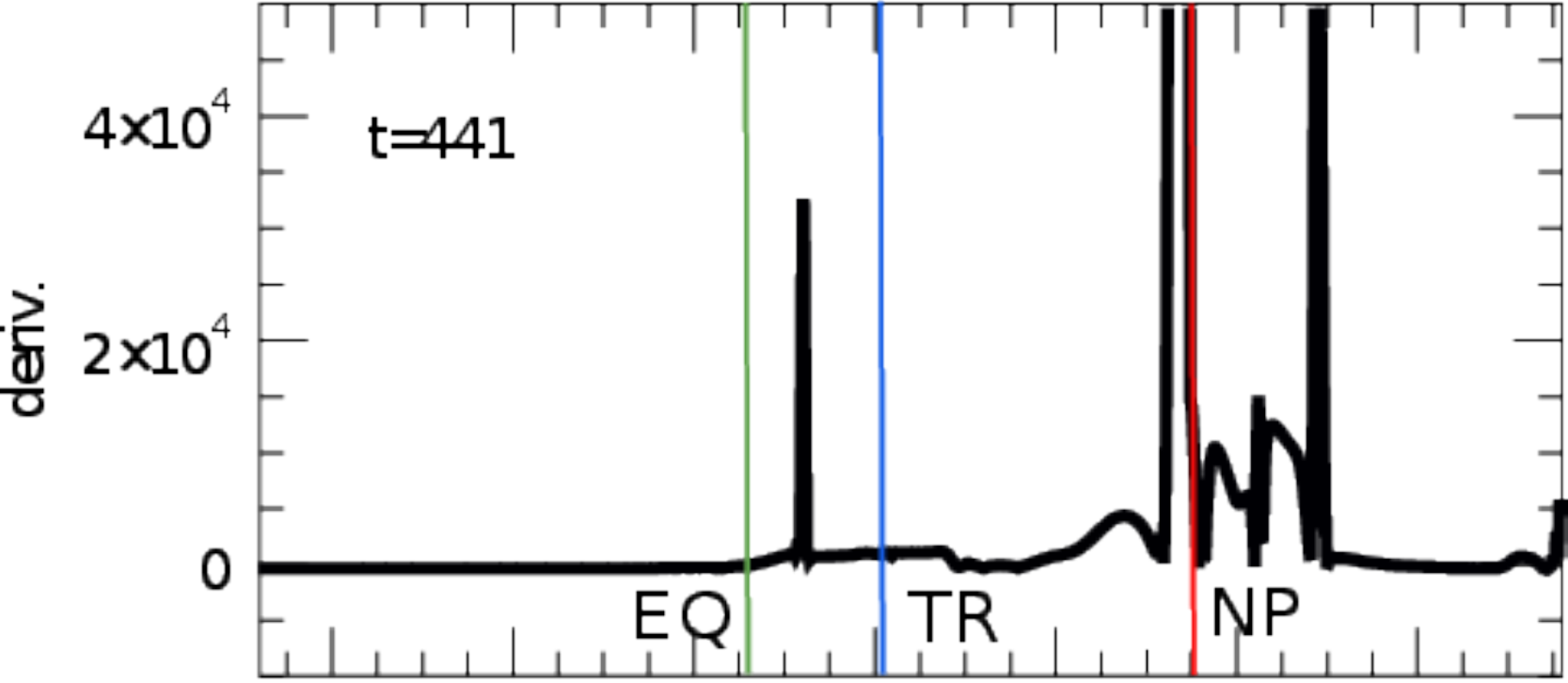} 
\\
\includegraphics[width=6cm]{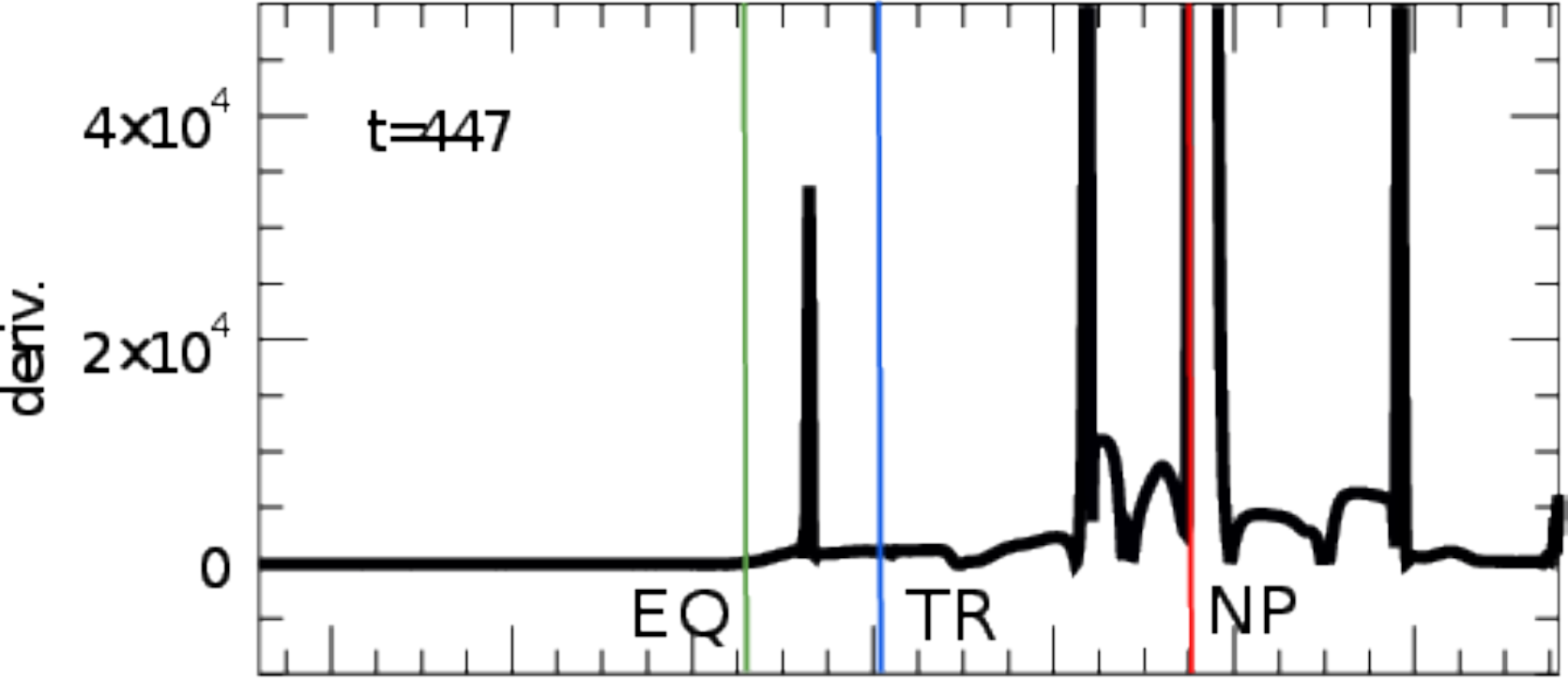} 
\\
\includegraphics[width=6cm]{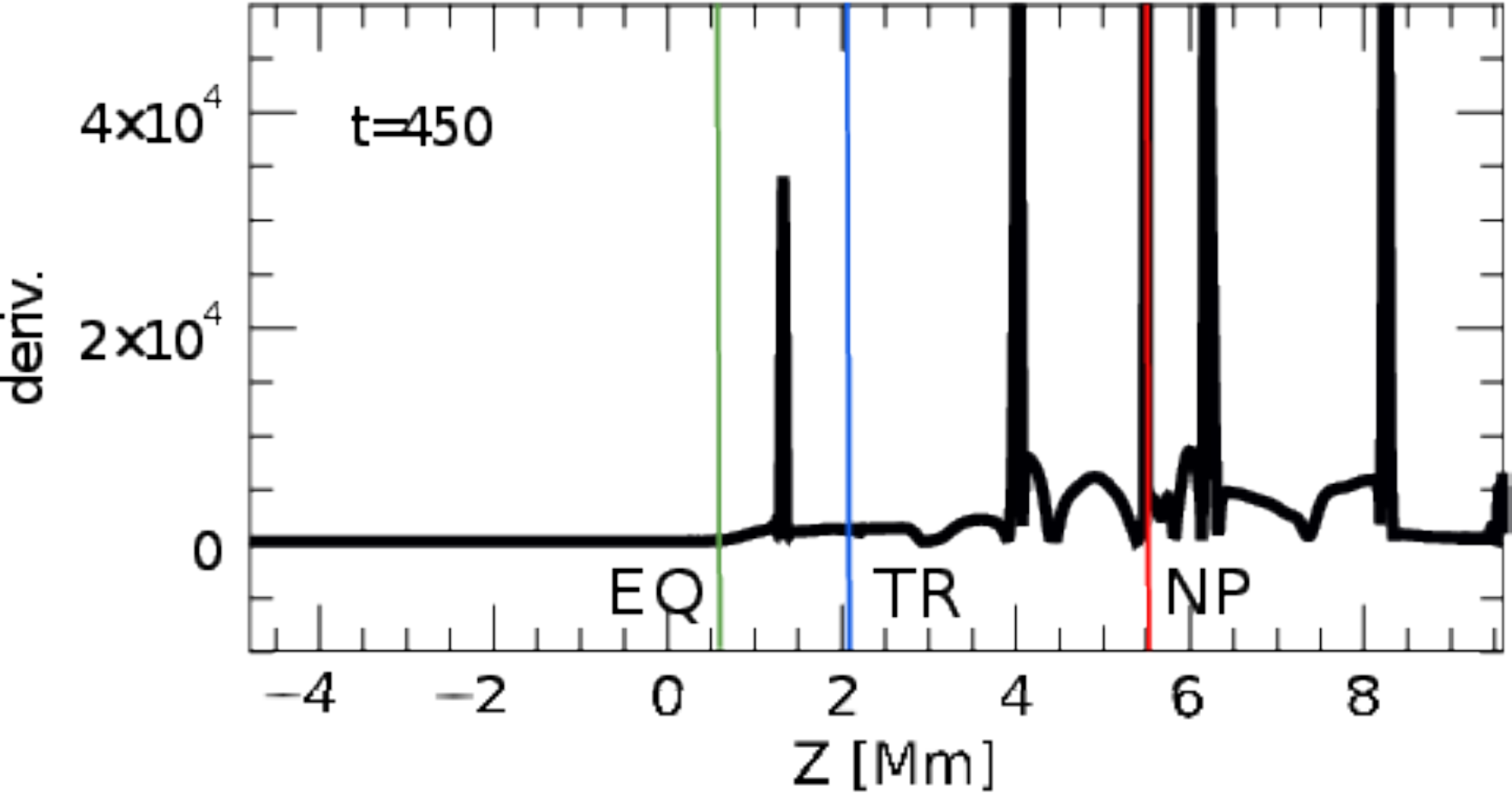} 
\caption{Temporal evolution of a slow magneto-acoustic shock wave developed below the null point (see the movie provided in the online version). In order to follow the discontinuity, the directional derivative of the velocity along the field lines (in $s^{-1}$) is shown. The locations of the transition region (TR), equipartition layer (EQ) and the null point (NP) are marked by vertical colored lines.} \label{fig:shock_NP}
\end{figure}

Figure \ref{fig:shock_NP} shows two shock waves developed at the same time in the chromosphere (black lines). Instead of showing the velocity evolution, its directional derivative along the field lines is plotted to facilitate the tracking of the discontinuities. The pair of shocks start propagating upwards. When the first shock reaches the null point at t=432 s (vertical line marked with NP), other shocks are formed at t=441 s and later. We called them ''secondary shocks''. Some of these shocks start propagating upwards and others propagate downwards (see the provided movie attached to Fig. \ref{fig:shock_NP}). Since the secondary shocks are of acoustic origin, they cause pressure variations in the direction parallel to the magnetic field. If we compare the time intervals when the secondary shock waves are developed and when the jet-like phenomena occur (see Figure \ref{fig:vlvt_nullpoint}), one can realize that both phenomena are taking place simultaneously. Hence, the pressure variations seen in the left hand panels of Figure \ref{fig:vlvt_nullpoint} are related to the secondary slow magneto-acoustic shock waves. This simultaneity leads us to conclude that the jet-like phenomena may be driven by slow magneto-acoustic shock waves formed as a result of the interaction between a slow magneto-acoustic shock wave and the null point.

\section{Frequency distribution}
\begin{figure}
\centering
\includegraphics[width=8 cm]{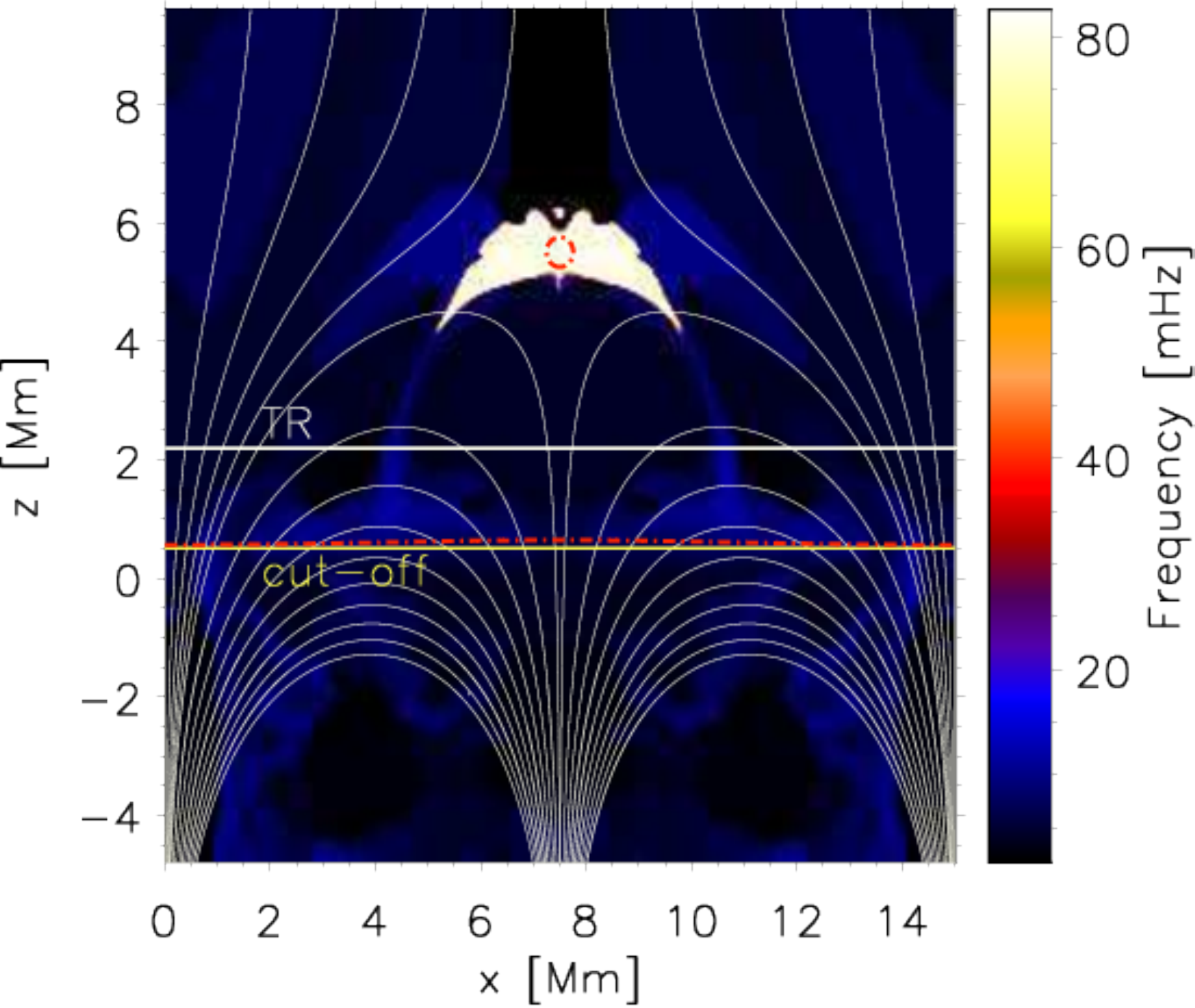}
\caption{Spatial distribution of the dominant frequency of the vertical velocity. The dashed-dotted red line marks $\beta=1$, and the solid lines mark the acoustic cut-off frequency location, transition region and magnetic field lines.} \label{fig:periods_nonlin}
\end{figure}

In their study, \citet{Santamaria+etal2016a} show a distinguishable region around the null point where high frequency waves dominate. This can be seen in Figure \ref{fig:periods_nonlin}, where the distribution of the dominant frequency of the oscillations in the simulation domain is shown. The high frequency region is clearly visible around the null point, while the oscillations are produced in the rest of the domain at frequencies below 20 mHz \citep[see Figure 3 in][]{Santamaria+etal2016a}. High frequency waves propagate further away from the null point, but their amplitude decreases with distance and lower frequencies start dominating. 

\begin{figure}
\centering
\includegraphics[width=8cm]{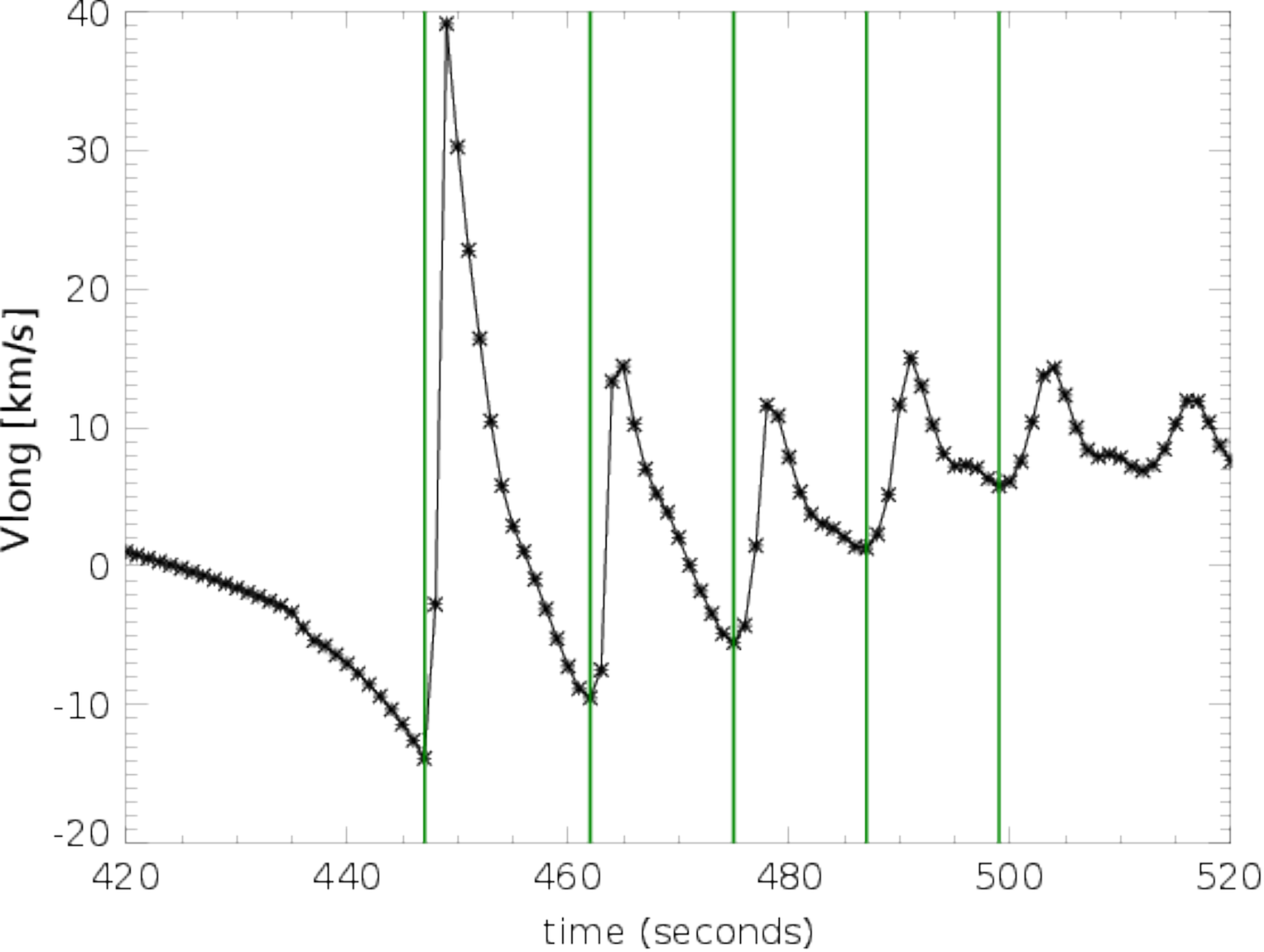}
\caption{Temporal evolution of the longitudinal velocity at a random location above the null point, for simulation times between 420 and 520 s. The vertical green lines mark the oscillation period. The asterisks represent the values of the velocity in each stored snapshot.} \label{fig:high_freq}
\end{figure}

To have a closer look at the high-frequency oscillations around the null-point,  Figure \ref{fig:high_freq} shows the temporal variation of the longitudinal velocity at a random location inside the high-frequency region. It covers a short period of time, between 420 and 520 seconds, when the strongest jet-like phenomenon takes place. One can observe strong supersonic non-linear oscillations with a period of 12-13 seconds (see the time intervals between the vertical green lines), corresponding to a frequency of 80 mHz. The points plotted on the curve mark the selected times when simulation snapshots have been stored (the time step in this simulation was of 2$\times$10$^{-3}$ s). Therefore, those high-frequency non-linear waves are well resolved in time in our simulations, excluding the possibility of being produced by numerical noise. Thus, we can conclude that the secondary shock waves are responsible for the high frequencies at the vicinity of the null point.

\section{Discussion and conclusions}
In this work we have carried out an analysis of the non-linear behavior of MHD waves around a two-dimensional null point. In particular, we have focused the study on the interaction between shock waves and the null point, which leads to a jet-like phenomenon that extremely alters the atmosphere around the null point. The jet-like feature is actually a train of secondary slow magneto-acoustic shock waves, formed by the interaction between a slow magneto-acoustic shock coming from the chromosphere and the null point. This phenomenon is pressure-driven, with the null point acting as a piston and driving waves outwards strongly along the field lines. The particular frequency of 80 mHz is the result of the combination of propagation speed of slow magneto-acoustic shocks, i.e. is a direct consequence of the thermodynamic structure of the atmosphere around the coronal null point. This suggests the possibility of recovering information about the state of the solar corona by detecting high frequency waves in complex magnetic field configurations. 

Shock wave propagation around null points was also modeled by \citet{McLaughlin+etal2009}. In their two-dimensional simulations, a null point was perturbed at its center. During the evolution of the system, jets were found to propagate outwards the null point and shock waves were generated. Contrary to our case, these authors found their shocks to be oblique magnetic shock waves. They initialized their simulation under the assumption of a cold plasma, which is heated after the shock development. Therefore, despite the qualitative similarity between the phenomena occurring at the null point with our case, the jets developed in their simulations are not driven by the same mechanism as the one we propose in the present paper. This way, our results complement those obtained by them.

Null points are theoretically predicted to be ubiquitous in the solar corona, and on the other hand, the acoustic waves can easily steepen into shocks. Therefore, the jet-like phenomenon developed in our simulation might be taking place frequently in the solar atmosphere. For instance, observations of flaring loops show that either long and short period waves propagate immediately after the flare development \citep{Nakariakov+Melnikov2009}. The short-period oscillations have periods between 8 and 12 seconds \citep{Melnikov+etal2005, VanDoorsselaere+etal2011} and they are interpreted as fast sausage modes. Taking into account that flares are developed where magnetic reconnection events can take place \citep{Shibata+Magara2011}, such as null points, and that the observed frequencies in those structures are very similar to those found around the null point in our simulation, we suggest that the observed short-period waves could also be interpreted as slow magneto-acoustic shock waves. The question to address now is whether these high frequency waves appear by other driving mechanisms or not. This question will be discussed more in depth in forthcoming works.

\begin{acknowledgements}
This work is partially supported by the Spanish Ministry of Science through projects AYA2010-18029 and AYA2011-24808.
This work contributes to the deliverables identified in FP7 European Research Council grant agreement 277829, ``Magnetic connectivity through the Solar Partially Ionized Atmosphere''.
\end{acknowledgements}

\aareferences

\end{document}